# Progressive Filtering Using Multiresolution Histograms for Query by Humming System

Trisiladevi C. Nagavi[1] and Nagappa U. Bhajantri[2]

[1]Department of Computer Science and Engineering S.J.College of Engineering
Mysore, Karnataka, India
`tnagavi@yahoo.com`
[2]Department of Computer Science and Engineering Government Engineering College
Chamarajanagar, Karnataka, India
`bhajan3nu@gmail.com`

**Abstract.** The rising availability of digital music stipulates effective categorization and retrieval methods. Real world scenarios are characterized by mammoth music collections through pertinent and non-pertinent songs with reference to the user input. The primary goal of the research work is to counter balance the perilous impact of non-relevant songs through Progressive Filtering (PF) for Query by Humming (QBH) system. PF is a technique of problem solving through reduced space. This paper presents the concept of PF and its efficient design based on Multi-Resolution Histograms (MRH) to accomplish searching in manifolds. Initially the entire music database is searched to obtain high recall rate and narrowed search space. Later steps accomplish slow search in the reduced periphery and achieve additional accuracy.
Experimentation on large music database using recursive programming substantiates the potential of the method. The outcome of proposed strategy glimpses that MRH effectively locate the patterns. Distances of MRH at lower level are the lower bounds of the distances at higher level, which guarantees evasion of false dismissals during PF. In due course, proposed method helps to strike a balance between efficiency and effectiveness. The system is scalable for large music retrieval systems and also data driven for performance optimization as an added advantage.
**Keywords:** Progressive Filtering, Multiresolution Histograms, Multiresolution Analysis and Query by Humming.

## 1 Introduction

Content based online music enabling systems are being developed and revamped in order to keep up with expectations of search and browse functionality. These approaches as a group describe the Music Information Retrieval (MIR) systems and have been the area under exhaustive research. The rationale of MIR research is to develop new theory and techniques for processing and searching music databases by its content. The QBH is a special branch of MIR and also a popular content based music retrieval method where the user enters a search query by humming.

Most of the research works on QBH[2][3][11][13][10] are based on the music processing and focused on components like melody extraction, representation, similarity measurement, size of databases, query and search algorithms. The strong literature supports the symbolic representation for melody in the form of zero-cross detection, energy, Modified Discrete Cosine Transform (MDCT) [5], pitch contour [10], rhythm [18] and quantized pitch change descriptor [11]. Also there is a remarkable amount of research work [6] [7] [12] in the broader areas of similarity measurement with reference to music patterns.

Most of the approaches proposed in the literature are not suited for real-world applications of music retrieval from a large music database. Perhaps, is due to either undue complexity in computation which leads to longer response time or performance degradation; subsequently leading to erroneous retrieval results. Striking a balance between computation and performance is the ultimate goal for such retrieval systems. As a result there are a few speeding up [14] [19] [20] mechanisms proposed for QBH.

Quite extensive literature[2][3][11][13][5][10][6][18][7][12] is available on QBH system, but there is no significant amount of literature[21][9][16][1][8] towards designing filtering procedures. Authors [9] have projected a mathematical analysis for a two-stage Query by Singing\Humming (QBSH) system, which is the first application of PF to QBSH. In another work authors [17] proposed the concept of iterative deepening Dynamic Time Warping (DTW), which is a special form of PF for speeding up DTW. Improvement in the form of multi phase PF for QBSH without much design analysis is presented in [19] [16]. Research work [17], proposes a simplified version of PF with a constant computation time with respect to survival rates for each comparison stage. However, most of the proposed methods still portray the deficit in meticulous investigation, efficiency and effectiveness.

Therefore, in this paper we have proposed to apply PF using MRH approach for QBH system to accomplish the improved retrieval accuracy. Real-world applications of music retrieval symbolize huge amount of non relevant songs with reference to user queries causing input imbalance problem. We expect that these two techniques are most applicable to mitigate the effect of input imbalance. The exhaustive experimentation substantiates the potential of proposed method to construct an effective music retrieval system based on humming input. In this paper, as explained above, we have motivated to use PF as a filtering procedure. So, the next section gives a brief view of PF used for search space reduction. In section 3 we have made diligent discussion on MRH framework for pattern matching in music retrieval systems. While section 4 elaborates the details on similarity measure stratagem for QBH. In section 5, experimental results are presented and discussed. The last section enumerates the conclusion.

## 2  Progressive Filtering (PF)

The inspiration behind PF is to apply a series of comparisons, in which each comparison will select a smaller set that is likely to contain the target of the input query. The process is repeated until final output contains list of songs with appropriate length, say 10 or 20. PF on QBH is performed by applying multiple stages of comparisons between a query and the songs in the database, using an increasingly more complicated recognition mechanism to the decreasing candidate pool. So that the correct song will remain in the final candidate pool with a maximum probability. Intuitively, the initial few stages are quick and impure such that the most unlikely songs in the database are eliminated. On the other hand, the last few stages are more sophisticated and time consuming such that the most likely songs are identified [9].

After each stage of PF, the number of surviving candidates in the candidate pool of the database becomes smaller, and the recognition technique turns into refined and effectual. The final output is the surviving candidate songs at the last stage. The multistage representation of PF is shown in Fig. 1, where there are m stages, corresponding to different comparison methods with varying complexity.

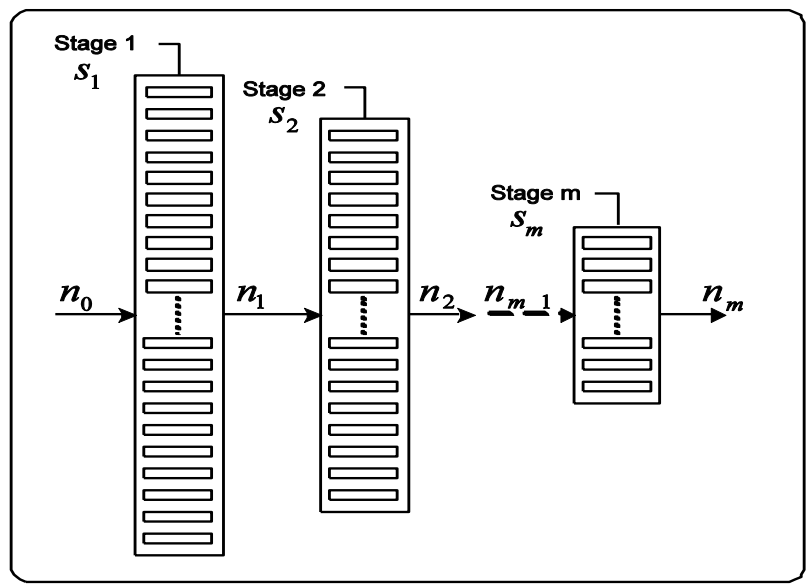

**Fig. 1.** Multistage Representation of Progressive Filtering

For stage $i$, the input is the query and $n_{i-1}$ surviving songs from the previous stage. The output of stage $i$ is a reduced set of candidate songs of size $n_i = n_{i-1} s_i$ for the succeeding stage $i+1$. In other words, each stage performs a filtering process that reduces the number of the candidate songs by a factor of the survival rate $s_i$. Each stage is characterized by its capability to select the most likely song candidates as the input to the succeeding stage. For a given stage, this capability can be represented by its recognition rate, which is defined as the probability that the target song of a given query is retained in the output song list of this stage. Intuitively, the recognition rate is a function of the survival rate.

## 3 Multi-Resolution Histograms (MRH)

### 3.1 Essence

Over the past few years Multi-Resolution Analysis (MRA) is receiving major attention by researchers in the domain of computer graphics, geometric modeling, signal analysis and visualization. It is a most important approach for proficiently representing signals at many levels of detail with numerous advantages like compression, different layers of details display and progressive transmission [4]. The term multi-resolution is used in diverse perspective such as multi-resolution based wavelets, subdivisions, hierarchies and multi-grids.

Histograms provide a very effective means of data reduction and depict many attributes of the data like location, spread, and symmetry. It is also possible to decompose music signal and build histograms on the underlying cumulative data distributions. Histograms give better approximation for cumulative data distributions with less space usage. However, histograms provide a comprehensive analysis of the data distribution by excluding sequence details of values. MRH depiction is proposed for enhanced discrimination of music data based on their position fine points to assist effectual QBH system. The music signal is recursively decomposed and cumulative histograms are built. Together all these cumulative histograms of a music signal are remarked as MRH. The selection of number of levels $l$ is directly proportional to precision. Early phase cumulative histograms exhibit lesser amount of music information than later phases. These early phase MRH are used to provide quick approximate answers to music retrieval queries in the beginning. Later phase of searching with next level MRH gives us better estimates.

In this paper, a MRH based representation is proposed to approximate music signal that is invariant to shifting and scaling. The MRH representation detects existence of a pattern along with shape matching. In the early phases of searching music signals with specific specified pattern are retrieved, then search continues for shape matching yielding result set of music signals that are of interest to the user. The hierarchical MRH framework is shown in Fig. 2. The symbol $HR_i$ indicates the histogram representation at level $i$.

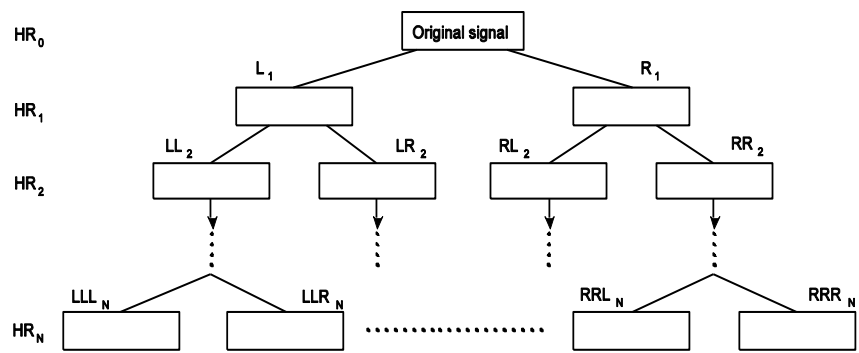

**Fig. 2.** Multi-Resolution Histogram Representation

## 3.2 Connotation of Mathematics

Histogram function $h_i$ counts the number of samples that fall into each of the disjoint sets known as bins. Thus, if $n$ is the total number of samples and $t$ is the total number of bins, the histogram function $h_i$ is defined as following:

$$n = \sum_{i=1}^{t} h_i \qquad (1)$$

A cumulative histogram function counts the cumulative number of samples in all of the bins up to the specified bin. In particular, the cumulative histogram function $hc_i$ of a histogram function $h_j$ is specified as:

$$hc_i = \sum_{j=1}^{i} h_j \qquad (2)$$

Cumulative frequency distributions authorize users to approximate frequencies over numerous bins. There is no standard value for number of bins, and different number of bins exhibit different features of the samples. Based on the data distribution and the objective of the analysis, different bin widths are chosen. The numbers of bins t are calculated from a recommended bin width $w$ as:

$$t = \left\lceil \frac{max(S) - min(S)}{w} \right\rceil \qquad (3)$$

where $S$=*samples to be histogrammed*. Also the equal sized bin widths are found by dividing the range with the number of bins $t$.

Primary objective of our research is to develop search criteria using similarity based queries over one dimensional music signal. Such music signal $S$ is defined as a sequence of values:

$$S = [s_1, s_2, \ldots s_N] \qquad (4)$$

where $N$, the number of samples in $S$ and $s_i$ is a vector of values that was sampled at timestamp $t_i$. Given a music signal database

$$D = \{S_1, S_2, \ldots S_M\} \qquad (5)$$

and a query $Q$, the aim is to find all the music signals in $D$ that contain the specified query $Q$ as well as histogram shape similar to that of $Q$. MRH are constructed by dividing the range $[min_D, max_D]$ of music database $D$ into $t$ non-overlapping equal size sub-regions, identified as histogram bins. Histogram $H_s$ is computed by counting the number of data values $h_i$ ($\leq i \leq t$) that are located in each histogram bin $i$.

$$H_s = [h_1, h_2, \ldots h_t] \qquad (6)$$

A cumulative MRH is a mapping that counts the cumulative number of observations in all of the bins up to the specified bin. That is, the cumulative histogram $HC_s$ of a histogram $H_s$ is defined as:

$$HC_s = \sum_{i=1}^{t} h_i \qquad (7)$$

MRH at higher levels have enhanced discrimination power; however, the computation of MRH Distance (MRHD) at higher scales is more expensive than those at lower levels. So the number of levels trade-off should be established to balance complexity and precision.

### 3.3 Proposed Strategy

MRH construction system for database D is depicted in Fig. 2 and steps are shown in the following algorithm 1.

**Algorithm 1:** Procedure to Construct Multi-Resolution Histograms for Music Database

**Input:** a music database $D$, number of levels $l$ and the number of histogram bins $t$
**Output:** a histogram data set $H_D$

1. level $l=0$
2. repeat
3. for each $S_i$ of database $D$ do
4.   divide the $S_i$ into $2^l$ non overlapping equal segments $S_{il,l}$ and $S_{ir,l}$
5.   locate $\max_D$ and $\min_D$ values of the $D$
6.   divide the range $[\min_D, \max_D]$ into t non-overlapping equal size bins $h_{il,l}$ and $h_{ir,l}$
7.   for each $S_{il,l}$ and $S_{ir,l}$ of $D$ do
8.     for each data point $s_{il,l}$ and $s_{ir,l}$ of $S_{il,l}$ and $S_{ir,l}$ respectively do
9.       for each bin $h_{il,l}$ and $h_{ir,l}$ do
10.         if $h_{il,lowerlimit} \leq s_{il,l} \leq h_{il,uppperlimit}$ then
11.           $h_{il,l} = h_{il,l} + 1$;
12.         end if
13.         if $h_{ir,lowerlimit} \leq s_{ir,l} \leq h_{ir,uppperlimit}$ then
14.           $h_{ir,l} = h_{ir,l} + 1$;
15.         end if
16.     end for
17.   end for
18.   end for
19. insert generated $H_{S_{il,l}}$ and $H_{S_{ir,l}}$ to the result data set $H_D$
20. end for
21. $l=l+1$ //increase level by 1//
22. until (l=user specified levels)
23. return the result data set $H_D$

## 4 Similarity Measure Stratagem for Query By Humming

### 4.1 Multi-Resolution Histograms Distance (MRHD) Measure

In order to recognize the query pattern in the music database, we have attempted to develop a similarity function which separately considers signal frequency as well as positional information. Given a song $S$ of music database $D$ and humming query $Q$, feature vectors $H_{S_f}$ extracted from song MRH are matched with query MRH $H_{Q_f}$ by means of the MRHD measure:

$$MRHD(H_{S_f}, H_{Q_f}) = \sum_{i=1}^{t} min(H_{S_i}, H_{Q_i}) \times \frac{\left(\sqrt{2} - d(H_{S_i}, H_{Q_i})\right)}{\sqrt{2}} \qquad (8)$$

where

$$d(H_{S_i}, H_{Q_i}) = \sqrt{\sum_{i=0}^{t} (h_{s_i} - h_{q_i})^2} \qquad (9)$$

is a Euclidean Distance function.

### 4.2 Database Pruning Using Threshold

MRHD for whole music database is calculated using equation 8 and 9. The average of the MRHD considered as the upper limit and 0 as the lower limit of threshold as shown in equation 10 and 11:

$$T_{upperlimit} = \frac{1}{M} \left( \sum_{i=1}^{M} MRHD(i) \right) \qquad (10)$$

and

$$T_{lowerlimit} = 0 \qquad (11)$$

where M=no of songs in the database. Unlikely songs are quickly eliminated by comparing MRHD values of database songs with threshold range. The song whose threshold is not in the range will be eliminated from the pruned database. In other words, if the following condition is not satisfied such song may be purged:

$$T_{lowerlimit} \leq MRHD_S \leq T_{upperlimit} \qquad (12)$$

This procedure is carried out at different histogram resolution level to form PF. The database pruning rate analysis is depicted in Fig. 4.

## 5 Results and Discussions

The relative performance of the proposed QBH method demonstrates several interesting trends and this section is dedicated to evaluate the proposed approach. Substantiation of feasibility of the proposed criteria is done through experimentation. In the sequel, three series of experiments were conducted with corresponding target and query corpus by varying the number of histogram bins from 100 to 1000 and histogram resolution level from 1 to 5. Finally, comprehensive discussions of performances are portrayed in terms of error rate, database pruning, Mean Reciprocal Rank (MRR), Mean of Accuracy (MoA) and Top X Hit Rate.

### 5.1 Target Corpus

We are proposing a novel QBH system exclusively for Indian music songs, so the corpus chosen for this study consists of 1000 Indian Kannada devotional monophonic MP3 songs. This collection is prepared from 39 subjects including songs from 22 males and 17 female singers. The corresponding training set includes a subset of 100, 200, 500 and 1000 songs for different experiments. MP3 songs contain convoluted melody information and even noise. Thus preprocessing is applied on the MP3 songs database to extract information needed by the system. In music, human vocal part always plays an important role in representing melody rather than its background music therefore it is desired to segregate both [15].

### 5.2 Query Corpus

For system evaluation, we employ a monophonic query corpus containing total 200 sample queries from ten participants. Each participant was asked to hum beginning of the target song two or three times each. The participants were selected from variety of musical backgrounds like with and without considerable musical training. Also they were instructed to hum each query as naturally as possible using the lyrics of the target corpus.

### 5.3 Error Rate Analysis

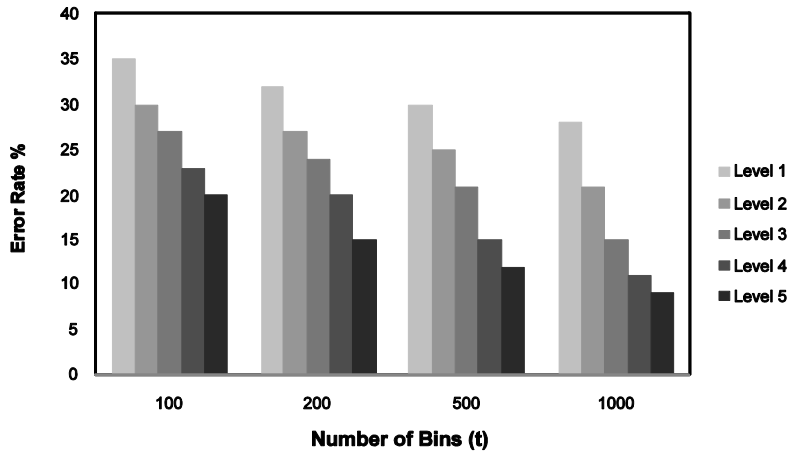

**Fig. 3.** Error Rate Analysis

Using the query and target corpus described above, the error rate is computed for the QBH system implementations presented in sections 2 through 4. Fig. 3 displays the error rate for five histogram resolution levels. The target database number of histogram bins is represented along the horizontal axis and the error rate along the vertical axis. As expected, direct comparison of error rates with increasing histogram bin numbers, yields the better performance, this improvement diminishes as the number of bins decrease.

Through prominent observation it was found that fine grain level music signal approximation is possible with higher number of histogram bins, which yields better performance. However, error rate increases with the decrease in the number of histogram bins.

### 5.4  Database Pruning Rate Analysis

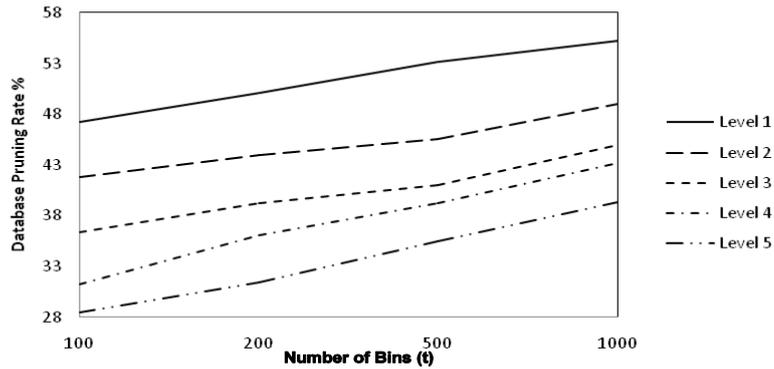

**Fig. 4.** Database Pruning Rate Analysis

Fig. 4 displays the pruning rate analysis for QBH system across different sized target databases with five histogram resolution levels. The target database's number of bins are represented along the horizontal axis and the pruning rate along the vertical axis. In this figure, the pruning rate for histogram resolution level 1,2,3,4 and 5 are shown with a line, dashed line, small dashed line, dash-dot line and dash-dot-dot line respectively.

Indeed, for increasing number of histogram bins and histogram resolution levels pruning rate is approximately 55% as shown in Fig. 4. The first histogram resolution level representation yields the most robust performance of pruning around 55%. For the target database with increasing number of histogram bins the best pruning rate is in the range 55.21% to 39.35% across different histogram resolution levels. That is, the histogram representation with higher number of histogram bins yields good    pruning rate, however it is computationally domineering.

## 5.5 Performance Analysis

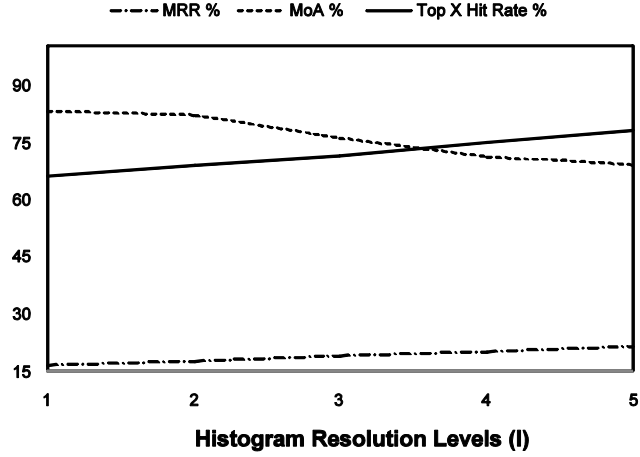

**Fig. 5.** Performance Analysis

Many different measures for evaluating the performance of QBH systems have been proposed [14] [9] [8]. The measures require a collection of training and testing samples for each test scenario and parameter combinations. The Mean Reciprocal Rank *(MRR)* is defined as:

$$MRR = \frac{1}{n}\sum_{i=1}^{n}\frac{1}{rank(t_i)} \quad (13)$$

MRR is a metric for estimating any system that generates list of potential responses to a query. Reciprocal rank of a query outcome is the multiplicative inverse of the rank of the first accurate response. That is, the MRR is estimated as the average of the reciprocal ranks of outcomes for a sample of queries. The reciprocal value of the MRR refers to the harmonic mean of the ranks. In other words frequency of the system estimating one of the first ranks is calculated through MRR [15]. We obtained MRR in the range 16.41% to 21.34% for different histogram resolution levels. The proposed strategy reveals that the MRR increases with increase in histogram resolution level as portrayed in Fig. 5. In other words, frequency of occupying top five ranks increases as histogram resolution level increases.

Similarly for each test scenario and parameter combination the Mean of Accuracy *(MoA)* is defined as:

$$MoA = \frac{1}{n}\sum_{i=1}^{n}\frac{n-rank(t_i)}{n-1} \quad (14)$$

It demonstrates the average rank at which the target was found for each query. We obtained MoA in the range 68.84% to 83.21% with histogram resolution levels one to five. From Fig. 5, it is found that the MoA decreases with increase in histogram resolution level. This indicates average rank of the retrieved song decreases with higher histogram resolution levels.

The Top X Hit Rate is defined as percentage of successful queries and it can be shown mathematically as:

$$Top(X) = \#\{rank(i): rank(i) \leq X\}/N \qquad (15)$$

where X symbolize top most songs and N indicates total number of songs. The impact of Top X Hit Rate for different histogram resolution level is portrayed in Fig. 5. The top X Hit Rate varied from 65.78% to 78.90% for different histogram resolution levels. From the Fig. 5, X value 10 was found to be the best, at which system obtained retrieval accuracy in the range 65.78% to 78.90 with increasing histogram resolution level.

Comparing figures 3, 4 and 5, the MRH based representations empirically yield relatively better performance in terms of MRR, MoA and Top X Hit Rate.

## 6 Conclusion

In this work, we have attempted to exploit advantages of MRA technique to progressively reduce search space for QBH applications. In these kinds of applications, initial result set consists of songs that have some specific patterns; subsequent steps perform relatively slow search in the small space to retrieve all songs whose histogram shape matches with query. MRH analysis is employed as database filtering procedure to support iterative search in the database to produce effective music retrievals. The results obtained from exhaustive experimentation are encouraging. Exhaustive exploration of the possibility of combining equal area bin histogram and MRA is to be considered as part of further investigation.